\documentclass{article}
\usepackage{float}
\usepackage{graphicx} 
\usepackage[numbers]{natbib} 
\usepackage[utf8]{inputenc}
\usepackage[dvipsnames]{xcolor}
\usepackage{amsmath}
\usepackage{amssymb}
\usepackage{parskip}
\usepackage[margin=1.1in]{geometry}
\usepackage{hyperref}
\hypersetup{
  colorlinks=true,
  urlcolor=blue,
  citecolor=blue
}

\title{\textbf{AI Developments for T and B Cell Receptor Modeling and Therapeutic Design}}

\author{
    Linhui Xie\textsuperscript{1,*},
    Aur\'elien P\'elissier\textsuperscript{2,*},
    Yanjun Shao\textsuperscript{1},
    Mar\'ia Rodr\'iguez Mart\'inez\textsuperscript{1,$\dagger$}
}

\date{\vspace{-0.5cm}}

\begin{document}

\maketitle

\textsuperscript{1}Department of Biomedical Informatics and Data Science, Yale School of Medicine, USA\\
\textsuperscript{2} Z\"urich University of Applied Sciences (ZHAW), Switzerland \\
\textsuperscript{$\dagger$}Corresponding author: \texttt{maria.rodriguezmartinez@yale.edu}\\
\textsuperscript{*}A.P. contributed equally with L.X.

\section*{\center{Abstract}}

Artificial intelligence (AI) is accelerating progress in modeling T and B cell receptors by enabling predictive and generative frameworks grounded in sequence data and immune context. This chapter surveys recent advances in the use of protein language models, machine learning, and multimodal integration for immune receptor modeling. We highlight emerging strategies to leverage single-cell and repertoire-scale datasets, and optimize immune receptor candidates for therapeutic design. These developments point toward a new generation of data-efficient, generalizable, and clinically relevant models that better capture the diversity and complexity of adaptive immunity.\\

\section{Introduction: AI for immune receptors in immunoepidemiology}

AI and machine learning (ML) are rapidly transforming our ability to read and design adaptive immune receptors. T cell receptors (TCRs) and B cell receptors (BCRs, whose secreted forms are antibodies) are central to immune recognition and, at the population level, their repertoires encode collective records of infection, vaccination, and tumor–immune interactions. With the increasing availability of high-throughput repertoire sequencing~\citep{friedensohn2017advanced}, these records can now be analyzed at scale. Meanwhile, advances in protein sequence modeling, structure prediction, and generative design have created powerful new tools for interpreting and engineering these receptors~\citep{hie2024efficient}. Together, these developments open new avenues in both immunoepidemiology and therapeutic discovery.

Adaptive immune repertoires are generated through V(D)J recombination and shaped by clonal selection~\citep{tonegawa1983somatic}. As individuals encounter pathogens or vaccines, specific clones expand and persist, leaving signatures in the circulating receptor pool. Adaptive immune receptor repertoire sequencing (AIRR-seq) applies deep sequencing to rearranged TCR and BCR genes, capturing millions of receptor sequences from a blood or tissue sample~\citep{robins2009comprehensive, boyd2009measurement}. Here, a clonotype is defined by a unique V(D)J rearrangement; these clonotypes can be grouped into clonal lineages (i.e., clonotype-derived families, especially for B cells where somatic hypermutation diversifies sequences), allowing researchers to track how specific clones expand, contract, or persist after infection or vaccination and, in turn, to read out past and ongoing immune exposures from the repertoire itself.

Several public resources support population-scale repertoire analysis. For BCRs, the Observed Antibody Space (OAS) database compiles large collections of antibody sequences across many studies and conditions~\citep{olsen2022observed}. For TCRs, the Observed TCR Space (OTS) provides an analogous resource for T cell receptor repertoires~\citep{raybould2024observed}. Epitope annotations and reference information can be obtained from the Immune Epitope Database (IEDB)~\citep{vita2025immune}. In addition, the AIRR Data Commons~\citep{christley2020adc} and the iReceptor Gateway~\citep{corrie2018ireceptor} provide infrastructure for standardized data formats and metadata, enabling large-scale comparisons across cohorts for both TCRs and BCRs. These resources support key questions in immunoepidemiology, such as which epitope-specific motifs recur across individuals, how stable such motifs are over time, and how much observed convergence arises from shared exposures versus shared human leukocyte antigen (HLA) backgrounds.

Despite their richness, immune repertoires present analytical challenges. The sequence space is extremely large, labels (e.g., antigen specificity, binding affinity), are sparse and noisy, and informative patterns may be subtle or context-dependent. Classical approaches using k-mer frequencies \cite{glanville2017identifying}, motif mining \cite{ostmeyer2019biophysicochemical}, or repertoire-level summary features \cite{greiff2015bioinformatic} provide useful baselines but often fail to generalize across pathogens, sequencing platforms, or populations \cite{miho2018computational}.

AI models offer several advantages. First, recent works have applied self-supervised learning approaches and transformer models to large collections of unlabeled amino acid sequences, resulting in the development of protein language models (PLMs) \cite{rives2021biological, lin2023evolutionary}, which capture biological and functional properties, as well as secondary and tertiary protein structures directly from sequences. The number of PLMs is rapidly increasing, and some of them are specifically trained on TCR \cite{wu2021tcr, raybould2024observed} and BCR \cite{leem2022deciphering, kenlay2024large, olsen2022ablang, olsen2024addressing} sequences. 
These models can improve downstream tasks including paratope prediction, clonotype clustering, and repertoire comparison. A recent review of TCR–epitope binding prediction models is available~\citep{weber2024t}.

AI has also advanced structural modeling. Tools like AlphaFold3 \cite{abramson2024accurate}, IgFold \cite{ruffolo2023fast}, TCRmodel2 \cite{yin2023tcrmodel2} and TCRdock \cite{bradley2023structure} routinely predict receptor and complex structures, supporting structure-guided analyses of specificity and flexibility. More recently, generative models such as RFdiffusion \cite{watson2023novo,ahern2025atom} have enabled conditional protein design, accelerating discovery pipelines beyond traditional screening or directed evolution.

This chapter focuses on AI approaches for modeling, predicting, and designing TCRs and BCRs, emphasizing their applications in immunoepidemiology and immune receptor engineering. We begin with sequence-based models, particularly protein language models, which are well aligned with available data and widely usable. We then briefly discuss structure-based modeling and conclude with emerging generative models and integrative pipelines linking computation to experimental design and clinical translation.

\section{Data resources and computational challenges}
The development of AI models for TCR and BCR analysis depends on the availability of large, reliable datasets. Below, we summarize key public resources available for antibody and TCR repertoire analysis, spanning sequence, epitope, and structural information that together support studies of repertoire diversity and antigen recognition.

\paragraph{BCRs / Antibodies}
Large-scale sequence resources provide the foundation for antibody modeling and engineering. The Observed Antibody Space (OAS)~\citep{olsen2022observed} is the most comprehensive, containing over a billion unique antibody sequences aggregated from over 80 studies. Most sequences originate from human and mouse repertoires and include both heavy and light chains; while the majority are unpaired, OAS also provides paired variable heavy (VH) and  variable light (VL) sequences ($\sim$150k). By capturing the diversity of natural repertoires in a standardized, annotated format, OAS has become a cornerstone for pretraining antibody-specific language models. Complementing this, community-driven repositories such as the AIRR Data Commons~\citep{rubelt2017adaptive} and iReceptor~\citep{corrie2018ireceptor} enable standardized access to repertoire data from diverse studies, thereby supporting reproducibility and large-scale integrative analyses.

In addition to general repertoire collections, several specialized datasets focus on therapeutic or antigen-specific contexts. Thera-SAbDab catalogs clinical-stage and approved therapeutic antibodies and links sequences to their molecular targets, which is valuable for drug discovery~\citep{raybould2020thera}. The recently introduced antibody-specific epitope prediction (AsEP) database organizes experimentally validated antibody–epitope pairs and provides a benchmark for epitope prediction~\citep{liu2024asep}. Functionally annotated sequences remain relatively sparse overall, with the densest labels around viral pathogens. For SARS-CoV-2, resources such as CoV-AbDab~\citep{raybould2021cov}, Ab-CoV~\citep{rawat2022ab}, and AlphaSeq~\citep{engelhart2022dataset} aggregate sequences with binding or neutralization readouts, especially for antibodies targeting the receptor binding domain of spike. For human immunodeficiency virus (HIV), CATNAP (Compile, Analyze and Tally NAb Panels) curates broadly neutralizing antibodies together with breadth and potency across viral isolates~\citep{yoon2015catnap}. Deep mutational scanning of SARS-CoV-2 receptor-binding domain (RBD) adds residue-level maps that link specific substitutions to binding and escape phenotypes~\citep{greaney2022antibody}.

For binding and affinity readouts, several curated resources provide quantitative labels. Structural kinetic and energetic database of mutant protein interactions (SKEMPI)~\citep{jankauskaite2019skempi} remains a primary reference for studying mutation effects in antibody–antigen complexes, reporting relative changes in binding free energy ($\Delta\Delta G$). Binding free energy ($\Delta G$) quantifies the strength of the interaction between an antibody and its antigen, with larger negative values indicating stronger, more favorable binding. The notation $\Delta\Delta G$ (``delta-delta-G'') represents the change in binding free energy caused by a mutation. Notably, SKEMPI covers a wide range of targets across both viral and non-viral systems, with measurements including viral proteins (Sars, HIV, influenza) as well as human proteins. More recent compilations broaden the landscape, including AbDesign~\citep{janusz2025abdesign} and AbBiBench~\citep{zhao2025benchmark}. In parallel, SAbDab~\citep{dunbar2014sabdab} aggregates direct affinity measurements, including dissociation constants ($K_d$), for a subset of antibody–antigen complexes. The $K_d$ measures how tightly an antibody binds to its antigen, representing the concentration at which half of the antibody molecules are bound. Lower $K_d$ values (e.g., nanomolar or picomolar) indicate stronger, higher-affinity binding, while higher values indicate weaker interactions. While the number of experimentally labeled entries remains in the low thousands, recent studies have supplemented these resources with synthetic labels generated by binding predictors, scaling coverage to hundreds of thousands of examples and enhancing downstream model performance despite their synthetic nature~\citep{wu2025simple}. Despite rapid progress, functional datasets remain heterogeneous in both format and scope, with outcomes reported as dissociation constants ($K_d$), inhibitory concentrations (IC\textsubscript{50}), escape fractions, or model-derived scores. This variability complicates direct comparison across studies and underscores the need for unified functional benchmarks that can maximize the impact of AI-driven antibody discovery~\citep{liu2025abrank}.

\paragraph{TCRs}
For TCR sequence analysis, several well-known databases provide the training foundation for models that try to predict antigen specificity. VDJdb is a central resource that offers a curated collection of TCR sequences paired with viral or self-antigen targets~\citep{shugay2018vdjdb}. This makes it useful for supervised machine learning, where models are trained on examples with known outcomes. At the same time, annotation quality varies and labels can be noisy: VDJdb aggregates evidence across diverse experimental pipelines, and false positives can arise (e.g., from non-specific pMHC multimer staining), which motivates confidence scoring and additional validation procedures such as functional assays~\citep{shugay2018vdjdb}. Consistent with this, a recent large-scale independent functional re-testing of a substantial fraction of VDJdb-deposited TCR--pMHC pairs reported that claimed TCR reactivity was confirmed for only \(\sim\)50\% of evaluated entries and released a validated subset (TCRvdb), highlighting substantial label noise in historical training data~\citep{messemaker2025functionally}. More broadly, “positive” TCR--antigen labels can also mix different measurement types---from binding-based readouts (e.g., multimer binding or biophysical affinity measurements such as SPR) to downstream functional activation assays---so labels may not correspond to a single biological endpoint across studies~\citep{feng2024sliding}. McPAS-TCR complements this by linking TCR sequences with human pathogens, cancers, and autoimmune diseases~\citep{tickotsky2017mcpas}. Whereas VDJdb focuses on epitopes, McPAS emphasizes clinical context. More recently, TRAIT has sought to unify and expand these efforts by providing a comprehensive database of TCR--antigen interactions with standardized annotations across receptors, epitopes, and experimental conditions~\citep{wei2025trait}. The Observed T-cell Receptor Space (OTS) database further broadens coverage by systematically cataloging TCRs observed across many repertoires together with their basic annotations, offering a large but weakly labeled landscape of TCR usage~\citep{raybould2024observed}. ImmuneCODE shows the value of rapid data sharing~\citep{nolan2025large}. Released during the COVID-19 pandemic, it provided millions of TCR sequences responsive to SARS-CoV-2. This accelerated research but also showed the risks of fast data release, including uneven annotation quality and limited metadata. Complementing these sequence-centric resources, BATCAVE aggregates deep mutational scanning data for peptide variants, linking systematic peptide mutations to functional readouts~\citep{banerjee2025t}. Together with structure-focused databases such as TCR3d 2.0~\citep{lin2025tcr3d} and ATLAS~\citep{borrman2017atlas}, which links binding affinities with structures for wild-type and mutant TCR--pMHC complexes, these resources provide mechanistic constraints that connect sequence, structure, and biophysical phenotype for model development. To manage and share the growing volume of immune repertoire data, the iReceptor platform set rules for data formatting and sharing~\citep{corrie2018ireceptor}, building on AIRR (adaptive immune receptor repertoire) community standards~\citep{rubelt2017adaptive}. iReceptor serves as a hub for raw and processed sequencing data across many studies. Most of these sequences do not carry functional labels, so extensive pre-processing is needed before they can be used in AI models.\\

\section{AI models for immune receptors}

\subsection{The pre-language model era}
Before the advent of transformer-based language models, most computational work on BCRs and TCRs relied on simple sequence encodings, motifs, k-mer frequencies, or one-hot representations, paired with classical classifiers such as SVMs (support vector machines), random forests, or logistic regression~\citep{jokinen2021predicting, weber2024t}. These approaches captured local sequence patterns but could not model long-range dependencies, limiting their ability to generalize.

Nevertheless, they were effective for specific tasks. Fpr antibodies, early paratope prediction tools such as Parapred~\citep{liberis2018parapred} leveraged motif enrichment and local sequence windows to identify binding residues concentrated in complementarity-determining regions (CDRs), which follow a compact vocabulary~\citep{akbar2021compact, robert2022unconstrained}. These methods also helped uncover convergent evolution in BCR repertoires, identifying antibodies from different lineages binding the same epitope~\citep{friedensohn2020convergent, pelissier2023convergent, wong2021ab}. Antibody affinity prediction saw some success using repertoire-level features~\citep{erlach2025antibody}. For TCRs, analogous methods were developed for epitope specificity prediction, often using k-mer–based features or alignment-derived motifs~\citep{glanville2017identifying, karnaukhov2024structure}. Repertoire-level classification tasks in both receptor types applied bag-of-words models, k-mer statistics, or alignment-free similarity metrics to detect disease-associated signatures, though performance was hampered by noise and sparsity.
 
Other problems were addressed more effectively with non-ML methods rooted in immunology and evolutionary biology. Clone assignment for both BCRs and TCRs typically relied on V and J gene usage combined with CDR3 similarity and clustering algorithms~\citep{richardson2021computational, pelissier2023convergent,pelissier2023exploring}, while phylogenetic inference drew on likelihood-based or Bayesian models that explicitly represent somatic hypermutation and clonal evolution~\citep{gupta2015change}.

In summary, early ML approaches across BCR and TCR modeling were driven by handcrafted features and shallow models. They performed reasonably well in tasks with strong biological priors, such as identifying CDR-based binding sites or grouping clonotypes, but fell short in open-ended problems like antigen binding prediction. These limitations set the stage for deep representation learning and the adoption of protein language models.

\subsection{BCR / antibody language models}

The rise of large-scale antibody sequencing datasets and curated structural repositories has fueled the development of AI models for antibody representation and design. These models span from sequence-based embeddings to structure prediction, affinity estimation, and generative design.

\paragraph{Antibody language models} 

\begin{figure}[ht]
\centering
\includegraphics[width=0.95\textwidth]{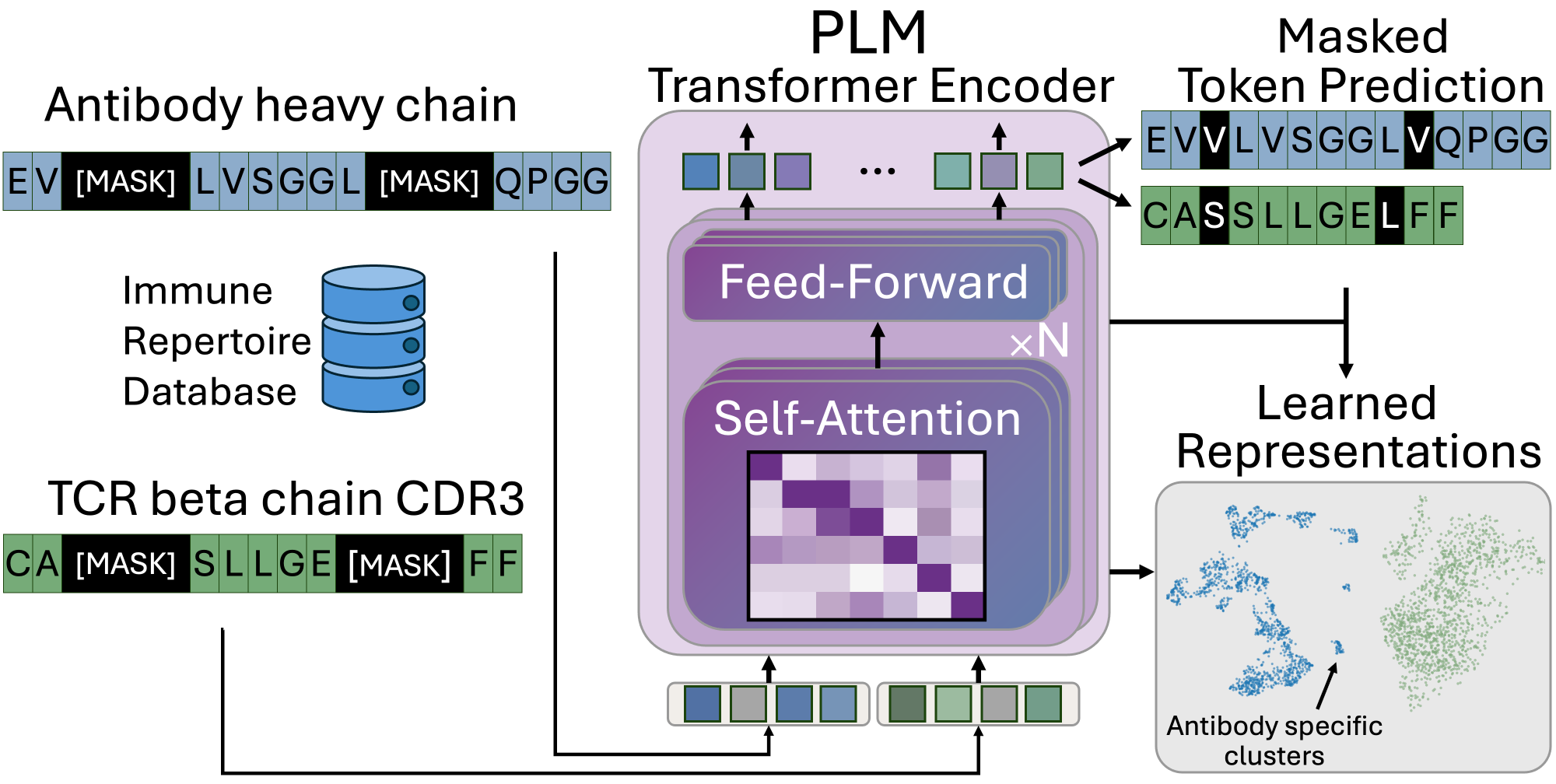}
\caption{Masked language modeling architecture for immune receptor representation. Antibody and TCR sequences are tokenized, with selected positions masked, and then processed through transformer encoder layers from a PLM that learn contextualized embeddings via self-attention. Through the neural network's feedforward modules, the embeddings are further processed for downstream analyses, such as visualization in a lower-dimensional space. Pre-training on large repertoire databases (OAS for antibodies; VDJdb, ImmuneCODE for TCRs) enables the model to predict masked residues and capture sequence patterns relevant to antigen recognition. The resulting embeddings support downstream tasks including binding prediction, specificity modeling, and repertoire analysis.
}
\label{fig:plm_architecture}
\end{figure}

PLMs have become central to antibody representation and design (Fig. \ref{fig:plm_architecture}). These models are inspired by bidirectional encoder representations from transformers (BERT)~\citep{devlin2019bert}, a breakthrough architecture originally developed for natural language processing that learns to understand text by predicting masked (hidden) words from their neighboring context. When adapted for proteins, BERT-style models treat amino acid sequences like sentences, learning the \emph{grammar} of protein structure and function by predicting masked residues.
Architecturally, these models use transformer encoders, which excel at capturing long-range dependencies in sequences. Each amino acid token is first converted into a numerical vector representation (embedding), then processed through stacked transformer blocks. Each block typically includes a self-attention layer—allowing each position in the sequence to “look at” every other position to determine which are most relevant—followed by a feed-forward layer that transforms the representations. This enables the model to weigh the importance of each amino acid relative to all others in the sequence. Through this architecture, PLMs such as ESM~\citep{rives2021biological}, ESM-2~\citep{lin2022language}, ProtBERT~\citep{brandes2022proteinbert}, aminoBERT~\citep{chowdhury2022single}, ProGen~\citep{madani2023large}, Saprot~\citep{su2025democratizing}, and XTrimoPGLM~\citep{chen2025xtrimopglm} are trained on hundreds of millions of protein sequences, treating amino acids as tokens of a biological language~\citep{bepler2021learning} (Fig. \ref{fig:plm_architecture}). 

PLMs capture broad structural and functional patterns, enabling them to predict protein folds, binding events, and functional sites~\citep{bepler2018learning, lin2023evolutionary, hou2023learning, yeung2023alignment}. The contextualized embeddings produced by these models provide powerful starting points for downstream tasks such as paratope identification, affinity estimation, or developability prediction, often reducing training time and improving accuracy. However, their considerable size (up to billions of parameters) makes them computationally demanding, and the largest variants may require specialized hardware clusters (e.g., GPUs) even for relatively simple tasks such as computing embeddings.

Recognizing the unique properties of antibody repertoires, antibody-specific PLMs have been developed~\citep{vu2024linguistics}, including AntiBERTy~\citep{ruffolo2021deciphering}, AntiBERT~\citep{leem2022deciphering}, AbLang~\citep{olsen2022ablang, olsen2024addressing}, and AbMAP~\citep{singh2023learning}. Trained directly on millions of immunoglobulin heavy and light chain sequences, these models learn repertoire-specific features such as VDJ gene usage, CDR loop diversity, and patterns introduced by somatic hypermutation. As a result, their embeddings can outperform general PLMs in antibody-specific tasks such as paratope prediction, or B~cell maturation stage prediction~\citep{wang2023pre}.

Two main strategies have emerged for applying PLMs to antibody modeling. One is to adapt a general-purpose protein model through fine-tuning on antibody repertoires, as in ESM-ft~\citep{burbach2024improving}, thereby combining broad protein-language priors with repertoire specialization. Fine-tuning is an additional training step on a specialized dataset to adapt a pretrained language model for a specific task. This can be achieved with fine-tuning or with parameter-efficient methods such as LoRA (low-rank adaptation)~\citep{hu2022lora}, which has been shown to improve task-specific performance compared with frozen embeddings~\citep{schmirler2024fine}. Interestingly, LoRA freezes most of the model and updates only a small number of newly introduced parameters, reducing computational cost.

The other strategy is to train models entirely on antibody repertoires, as in AbLang2~\citep{olsen2024addressing}, which is optimized for immune-specific sequence distributions. These approaches encode information differently~\citep{deutschmann2024domain, wang2023pre}: general PLMs tend to excel when structural knowledge is critical, while antibody-specific PLMs perform better in repertoire-driven applications. Still, because specialized models are trained on narrower datasets, there is a risk of missing general protein knowledge that supports generalization to unseen antigens or structural contexts~\citep{singh2023learning}.

\begin{figure}[ht]
\centering
\includegraphics[width=0.95\textwidth]{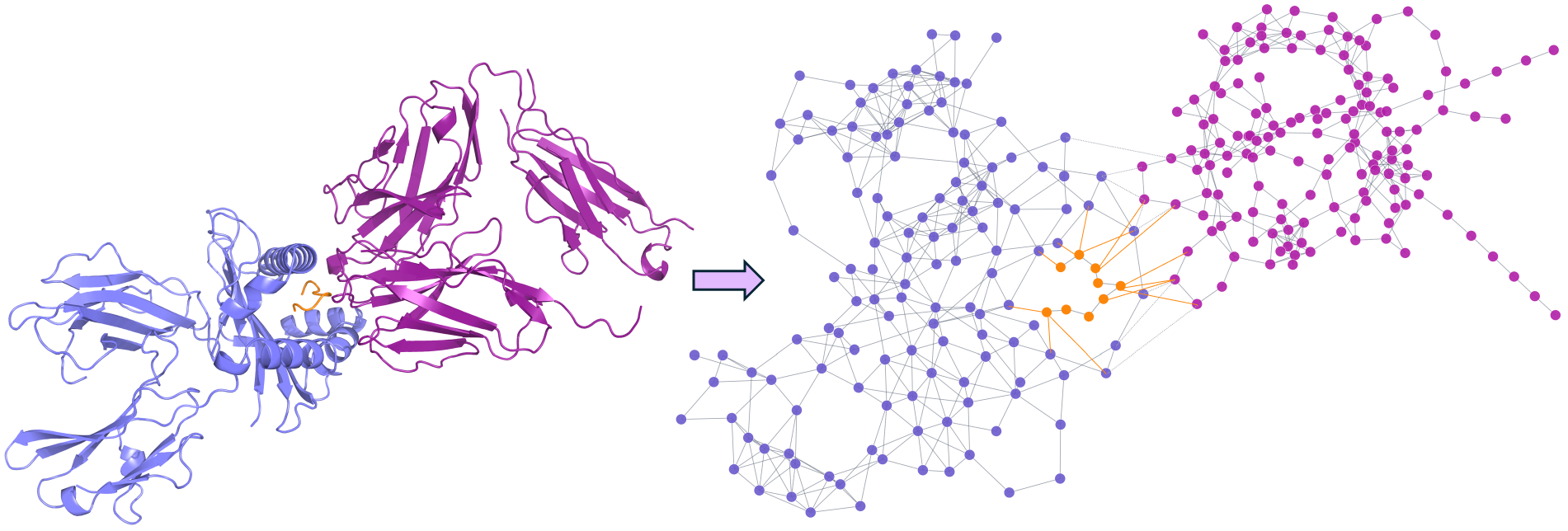}
\caption{Conversion of protein structures to graph representations for geometric deep learning using Graphein~\citep{jamasb2022graphein}. TCR-p-MHC complex structures (1AO7) from the Protein Data Bank (PDB)~\citep{berman2000protein} can be transformed into molecular graphs where nodes represent atoms or residues and edges encode spatial proximity or chemical bonds, enabling graph neural networks to learn interaction patterns.} 
\label{fig:tcrpmhc_to_graph}
\end{figure}

More recently, language models have begun to integrate both sequence and structural information, addressing limitations of sequence-only embeddings. To leverage structural data in machine learning, protein structures can be converted into graph representations (Fig.~\ref{fig:tcrpmhc_to_graph}), where nodes represent residues or atoms and edges encode spatial relationships, enabling geometric deep learning approaches. Notably, ESM-3~\citep{hayes2025simulating} jointly models amino acid sequences and atomic coordinates, enabling representations that capture both evolutionary and structural constraints within a single framework. In the antibody domain, IgBlend~\citep{malherbe2024igblend} extends this idea by combining repertoire-scale sequence data with experimentally determined and predicted 3D structures to build antibody-specific language models. Similarly, the immunoglobulin loop tokenizer (IGLOO) focuses on structural tokenization of CDR3 regions~\citep{fang2025tokenizing}, directly linking sequence variability with structural priors. Complementing these representation-centric approaches, ImmunoStruct~\citep{givechian2024immunostruct} adopts a multimodal fusion network that integrates peptide–MHC sequence, structural features, and biochemical properties to predict immunogenicity. These innovations promise more accurate paratope localization, improved affinity prediction, and ultimately stronger design capabilities than sequence-only approaches. More broadly in protein–protein interaction (PPI) modeling, analogous strategies enrich sequence embeddings with contact maps~\citep{guptanextgenplm} or incorporate known interaction networks such as MINT~\citep{ullanat2025learning} and PLM-interact~\citep{liu2025plm}, highlighting a growing trend toward multimodal representations that unify sequence and structure.

Together, these developments illustrate how PLMs have reshaped antibody informatics. General models provide a universal foundation, while antibody-specialized PLMs embed immune-specific features critical for therapeutic design. The choice between them depends on the task, balancing generalizability against repertoire-specific resolution.

\paragraph{Binding affinity prediction} Predicting binding affinity, and related tasks such as mutation impact or antibody (Ab) specificity, has long been a fundamental challenge in computational biology. State-of-the-art approaches typically adopt dual-encoder architectures~\citep{shan2022deep, jin2024attabseq, bandara2024deep, zhang2025antibinder, yuan2023dg}, where separate encoders process the Ab and antigen (Ag) sequences (or structures), and the resulting representations are fused via cross-attention or bidirectional attention mechanisms. These representations are then passed to a prediction head, often a convolutional neural network (CNN). More recently, since both Abs and Ags are proteins, graph-based representations have emerged as a natural way to capture structural and relational properties (Fig. \ref{fig:tcrpmhc_to_graph}). Tools such as Graphein~\citep{jamasb2022graphein} convert protein structures into molecular graphs, where nodes represent residues and edges encode spatial proximity or chemical interactions. Such models capture residue- or atom-level interactions through message-passing frameworks, with graph attention networks (GATs)~\citep{velivckovic2017graph} and graph convolutional networks (GCNs)~\citep{kipf2016semi} commonly used to model both local and global structural dependencies.

Regarding input data, earlier approaches relied solely on sequence information. However, there is growing consensus that incorporating 3D structural data of the Ab–Ag complex leads to substantial performance improvements. Recent studies have shown that structural and geometric features capture critical interaction details, resulting in significantly higher prediction accuracy~\citep{gong2024abcan, shan2022deep, cai2024pretrainable, myung2020mcsm, zhang2020mutabind2, tubiana2022scannet, zeng2023identifying}.

Models can also be categorized by their learning objectives. Most focus on predicting the effect of mutations on binding (e.g., GearBind~\citep{cai2024pretrainable} and Bind-ddG~\citep{shan2022deep}), while only a subset directly predicts binding affinity from the full Ab–Ag complex (e.g., ANTIPASTI~\citep{michalewicz2024antipasti}). This emphasis on mutation impact is driven by two factors: (i) the lack of a standardized framework for large-scale, general affinity prediction, and (ii) the importance of modeling local perturbations for applications such as affinity maturation and identifying viral escape mutations.

\paragraph{Other antibody properties} Beyond affinity, industrial discovery and development optimize a broader set of properties and attributes that determine whether a candidate can advance to the clinic~\citep{gordon2025therapeutic, erasmus2024aintibody}. Notably, reducing affinity has in some cases improved survival in cancer mouse models~\citep{yu2023reducing}, which highlights that it is only one of many parameters to balance. Key properties include immune reaction risk, expression yield, and developability metrics such as stability, solubility, viscosity, and manufacturability~\citep{gordon2025therapeutic, mason2021optimization}. 

Practical sequence-level triage often starts with tools like the therapeutic antibody profiler (TAP) to flag risk motifs and sequence liabilities~\citep{raybould2019five}. Solubility and aggregation risk are routinely estimated with sequence-based predictors such as CamSol~\citep{sormanni2015camsol} and Protein-Sol~\citep{hebditch2017protein}, and modern modeling pipelines assess surface features that correlate with poor manufacturability, for example patchy hydrophobic or highly charged regions linked to high viscosity~\citep{gordon2025therapeutic}.

An additional consideration is clinical safety, particularly immunogenicity. Immunogenicity is the likelihood that a therapeutic antibody elicits an unwanted immune response, often through anti-drug antibodies that reduce efficacy or cause adverse events. Because antibodies themselves can be processed by antigen-presenting cells, in silico risk estimation begins by predicting presentation of antibody-derived peptides on patient major histocompatibility complex (MHC) class II molecules using tools such as NetMHCIIpan and MARIA~\citep{reynisson2020netmhcpan, chen2019predicting}. Antibody-specific predictors then refine these scores by incorporating antibody context, for example whether a peptide lies in framework or CDR, its expected surface exposure and protease accessibility, and its similarity to common human germline or repertoire motifs~\citep{wang2024abimmpred, o2024modular, huang2025antibody}. This combination better prioritizes sequence edits for de-risking while preserving paratope function.

A complementary concept is humanness, which quantifies how closely an antibody sequence resembles natural human repertoires. Higher humanness generally correlates with lower immunogenicity risk, although it is not a guarantee. In practice, humanization and design workflows use repertoire-likeness (nativeness) scores to steer sequences toward human frameworks while preserving paratope function. Examples include Hu-mAb, a machine-learning approach that proposes humanizing mutations from large-scale human repertoire data~\citep{marks2021humanization}, and BioPhi, a platform that scores humanness and supports design and humanization decisions using repertoire-based and deep-learning models~\citep{prihoda2022biophi}. More recently, transformer-based nativeness metrics such as AbNatiV have improved calibration across species and germlines, providing more reliable guidance during hit selection and humanization~\citep{ramon2024assessing}.

Finally, chain pairing and repertoire assembly remain key constraints for library design and in silico selection. Recent learning-based matchers and large-scale repertoire resources improve heavy–light compatibility predictions beyond earlier co-occurrence heuristics~\citep{burbach2024improving, guo2025immunomatch}. End-to-end discovery systems increasingly treat these criteria jointly, using multi-objective optimization or constrained search to balance affinity with developability and safety, and to score candidates across the pipeline~\citep{erasmus2024aintibody, parkinson2025resp2}.

\begin{figure}[ht]
\centering
\includegraphics[width=0.9\textwidth]{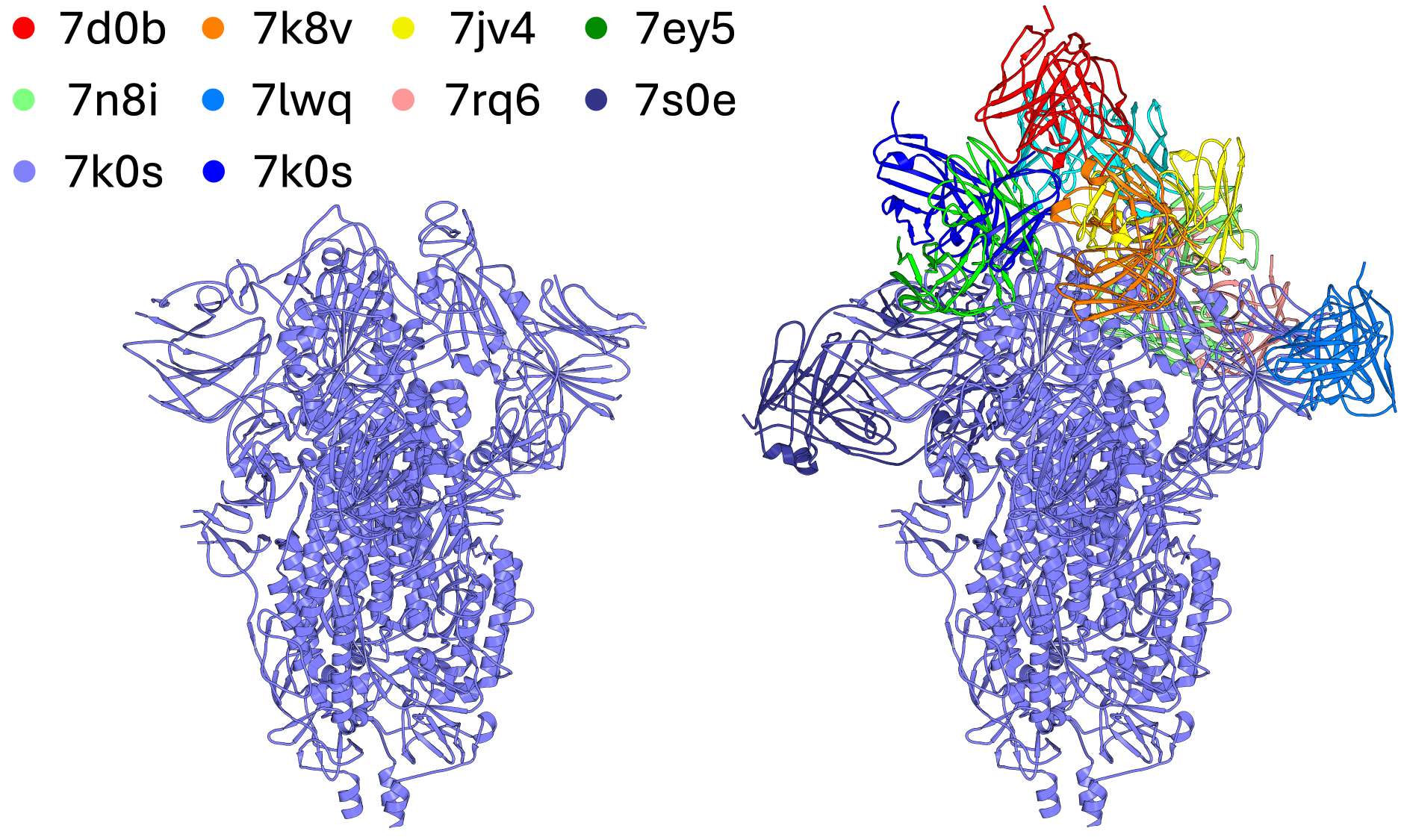}
\caption{Structural diversity of antibodies targeting the SARS-CoV-2 spike protein (light blue). (Left) unbound spike protein; (Right) multiple antibodies with distinct sequences (shown in rainbow colors) recognize different epitopes across the spike, including sites in the receptor-binding domain (RBD), N-terminal domain (NTD), and S2 stalk region, thereby contributing to the breadth of neutralizing immune responses.
}
\label{fig:covid_antibodies}
\end{figure}

\paragraph{Case studies}
The COVID-19 pandemic provided a clear proving ground for AI-guided antibody discovery and optimization using predictive models (Fig. \ref{fig:covid_antibodies}). 
A sequence–structure pipeline was used to propose and validate mutations to a human anti–SARS-CoV-2 antibody, achieving broader neutralization against emerging variants~\citep{shan2022deep}. In parallel, large antibody language models were used in predictive modes to score antigen–antibody pairing and prioritize candidate sequences for synthesis, enabling rapid in silico triage before wet lab work~\citep{he2024novo}.
Likelihood scores from general-purpose PLMs such as ESM can guide affinity maturation~\citep{hie2024efficient}. Without explicit antigen information, their predictive approach proposed evolutionarily plausible mutations that improved binding up to 160-fold, while also enhancing thermostability and viral neutralization. Complementing these design advances, EVEscape~\citep{thadani2023learning} showed that forecasting immune escape from historical sequence data is feasible at scale. Trained on sequences available before 2020, it anticipated a substantial fraction of SARS-CoV-2 mutations that later appeared, flagged escape at therapeutic antibody epitopes, and generalized beyond coronaviruses to influenza, HIV, Lassa, and Nipah.

Machine learning has also accelerated the discovery and refinement of broadly neutralizing antibodies across pathogens through predictive prioritization. 
For HIV-1, RAIN is a supervised repertoire-mining approach that identifies bnAb-like sequences, recovering known bnAbs and proposing new candidates for validation~\citep{foglierini2024rain}.
For influenza, a structure-based computational design framework improved the breadth and potency of a cross-subtype H5/H1 antibody through epitope–paratope analysis with experimental validation~\citep{duan2025computational}. Large-scale modeling of H5N1 evolution has mapped gaps in existing bnAb coverage, providing predictive guidance for redesign priorities and surveillance focus~\citep{ford2025large}. Together, these studies illustrate how antibody-focused PLMs, structure-aware predictors, mutation and developability scoring, and escape forecasting are translating into practical impact across pathogens through prediction and prioritization, rather than direct sequence generation.

\subsection{TCR language models}
TCRs pose challenges that differ from those of antibodies. They recognize peptides only in the context of an MHC molecule, and their specificity is shaped by subtle, context-dependent contacts as well as by HLA diversity across individuals. Although fewer high-quality structures are available than for antibodies, new language models are beginning to learn useful sequence representations and predict which peptide–MHC (pMHC) complexes a TCR might recognize.

One class of methods focuses on representation learning from receptor sequences. Analogous to antibody PLMs (Fig.~\ref{fig:plm_architecture}), TCR language models such as TCR-BERT~\citep{wu2021tcr}, TCRLang~\citep{raybould2024observed}, and TULIP~\citep{meynard2024tulip} learn embeddings of TCR $\alpha$ and $\beta$ chains that support repertoire comparison and downstream prediction with limited labeled data. This line of work has expanded to single-cell paired-chain datasets, where self-supervised objectives exploit TCR$\alpha$--TCR$\beta$ co-occurrence to learn better joint representations~\citep{goldner2024self}. Closely related approaches use contrastive learning to structure the latent space—bringing receptors (or receptor–epitope pairs) with shared specificity closer together and pushing unrelated ones apart. Examples include autocontrastive pretraining and epitope-conditioned contrastive objectives (e.g., SCEPTR~\citep{nagano2025contrastive}, TouCAN~\citep{pertseva2024tcr}, and EPACT~\citep{ zhang2024epitope}).

A second class of methods directly models TCR–pMHC interaction. TITAN encodes TCR and epitope sequences separately and uses attention to align them for binding/recognition prediction~\citep{weber2021titan}. TEIM (TCR–Epitope Interaction Modeling) similarly targets TCR–epitope recognition, but does so by explicitly modeling residue-level contacts between receptor and peptide representations~\citep{peng2023characterizing}. More recent work increasingly incorporates HLA context and applies multitask learning across peptide–MHC–TCR relationships to improve transfer beyond a single epitope or allele, as in PISTE~\citep{feng2024sliding} and UniPMT~\citep{zhao2025unified}. Other approaches aim to improve generalization, especially under “unseen epitope” evaluation, by combining contrastive objectives with lightweight prompting, e.g., LightCTL~\citep{ye2025lightctl}, or integrating sequence and structural signals, e.g., TRAP~\citep{ge2025trap}.

Across these approaches, performance is often constrained less by model capacity than by the quality of supervision. Positive training pairs are scarce, and many pipelines rely on synthetic “negatives” created by randomly mismatching receptors and peptides. These negatives are typically \emph{untested} rather than truly non-binding, which introduces substantial label noise. Cross-reactivity further blurs any strict positive/negative boundary, since TCRs can recognize multiple peptides and assay outcomes can vary with experimental context~\citep{sewell2012must, petrova2012cross, rius2018peptide}. As the number of proposed predictors has grown, systematic benchmarking has become essential; a 2025 Nature Methods assessment comparing 50 TCR–epitope predictors across many datasets highlights that reported accuracy can be highly dataset- and epitope-dependent~\citep{lu2025assessment}.

\begin{figure}[ht]
\centering
\includegraphics[width=0.9\textwidth]{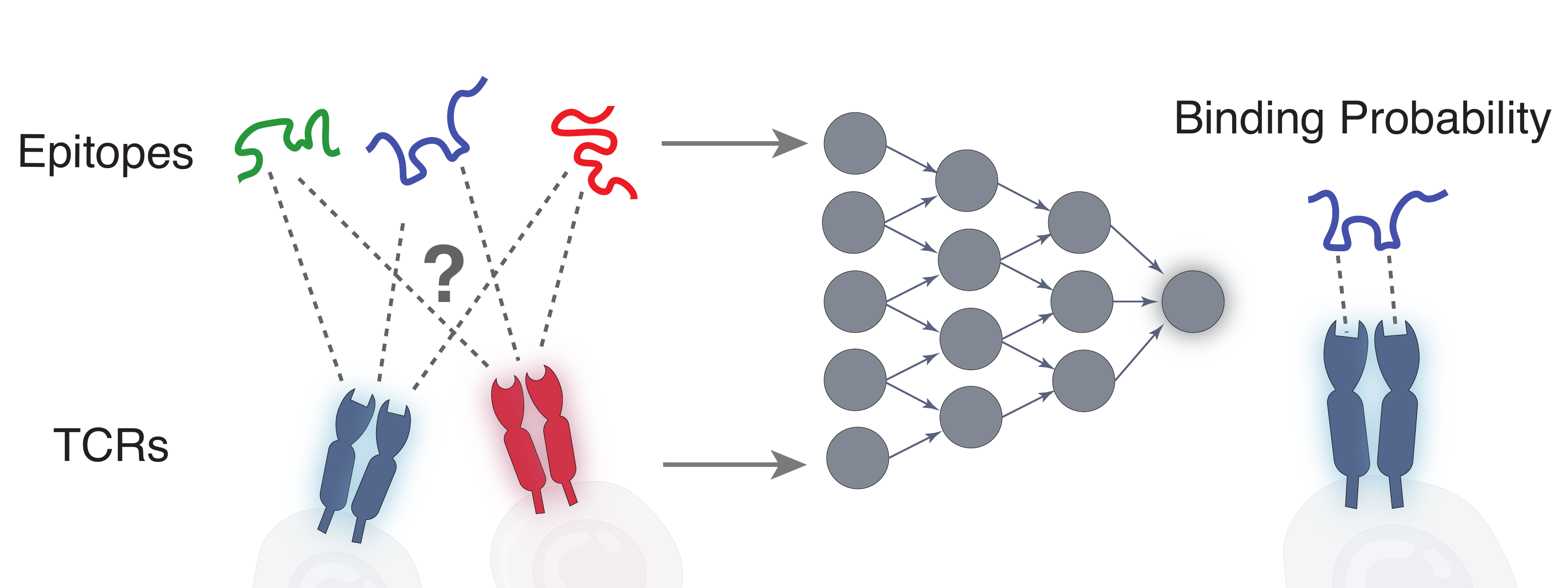}
\caption{AI framework for TCR-epitope specificity prediction using bimodal attention networks. TCRs and epitopes are encoded separately and then matched through attention mechanisms to predict binding probability, enabling repertoire-scale specificity screening.}
\label{fig:titan}
\end{figure}

\paragraph{Case studies} In cancer immunotherapy, researchers have used AI-guided screens to select or redesign TCRs that bind tumor neoantigens more tightly~\citep{li2023screening}, supporting adoptive T cell therapy. In infectious disease, sequence- and structure-aware predictors have helped map population responses to viral epitopes, including SARS-CoV-2~\citep{drost2024multi}. Together, these studies point to two practical payoffs. First, models can accelerate triage by narrowing large candidate lists to a manageable set for experimental validation. Second, models can flag potential cross-reactivity, informing risk assessment and vaccine design. In both settings, experimental checks remain essential.

\subsection{Current state of the field}
While antibody modeling has advanced rapidly thanks to the availability of massive repertoire datasets and thousands of solved antibody–antigen structures~\citep{dunbar2014sabdab}, TCR modeling remains comparatively less mature. Antibody-focused AI models benefit from the relative abundance of structural templates~\citep{abanades2023immunebuilder}, extensive high-throughput screening data, and clear developability benchmarks, enabling robust protein language models, structure predictors, and generative design pipelines. In contrast, TCR modeling is constrained by the limited paired TCR–pMHC data and structural coverage, and the inherent flexibility of CDR3 loops, which complicates prediction of binding geometry and specificity. As a result, antibody AI frameworks are already being translated into therapeutic discovery, while TCR models are still primarily in the exploratory and proof-of-concept phase~\citep{nielsen2024lessons}. Nonetheless, the rapid growth of TCR-focused datasets~\citep{lin2025tcr3d} and the development of multimodal~\citep{bradley2023structure}, structure-informed approaches suggest that this gap may narrow in the coming years.

\section{Generative AI for immune receptors engineering}
Generative models represent a new paradigm for immune receptor engineering. Rather than only predicting structure or affinity, these models aim to design new receptor sequences or structures with desired properties. Recent advances in deep generative modeling, particularly diffusion and flow-based approaches, have opened the door to controllable design of antibodies and TCRs. Diffusion learns to generate data by gradually removing noise from random starting points, and flow-based models learn smooth transformations between simple and complex distributions.

\subsection{Generative AI in Sequence Space} 

\begin{figure}[ht]
\centering
\includegraphics[width=0.9\textwidth]{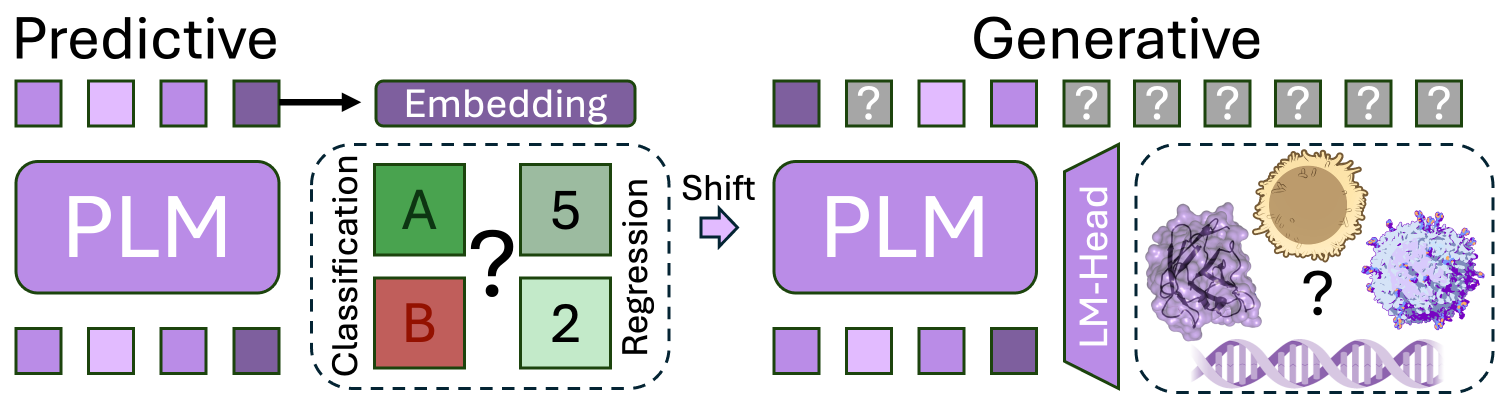}
\caption{Transition from predictive to generative language models in immune receptor design. (Predictive): A protein language model (PLM) encodes an input sequence into embeddings, which are then used for downstream tasks such as classification (e.g., binding vs.\ non-binding) or regression (e.g., affinity prediction). (Generative): The model is inverted, given desired properties or partial sequence constraints (indicated by ``?''), a PLM with a language modeling head (LM-Head) generates novel sequences predicted to exhibit target characteristics, enabling \textit{de novo} design of antibodies and TCR mimic binders for applications including viral neutralization, protein engineering, and therapeutic development.
}
\label{fig:predictive_to_generative}
\end{figure}

Generative models in sequence space offer a powerful paradigm for immune receptor engineering: rather than only predicting structure or affinity, these models aim to design new ones that satisfy desired functional, structural, or developability constraints (Fig.~\ref{fig:predictive_to_generative}). Four main approaches are emerging: (i) language model sequence generation, (ii) conditional or targeted generation, (iii) flow-based modeling of sequence ensembles, and (iv) reinforcement learning in sequence space.

\paragraph{Unsupervised protein language model for generation.}
Two main classes of language models are used for sequence generation. Masked language models (MLMs) are trained to predict hidden amino acids from their surrounding context, learning which residues are most probable at each position. Because these models score how well each amino acid fits its local sequence environment, they can prioritized mutations with the largest masked log-likelihood gains and observed that edits consistently favored by multiple independent protein LMs were enriched for affinity improvements~\citep{hie2024efficient}. Outside such settings, however, correlations between LM scores and functional outcomes (for example, affinity, stability) are often moderate and inconsistent~\citep{ucar2024exploring,van2024protein,zhao2025benchmark}. Moreover, MLMs primarily support local mutation rather than true de novo synthesis; while one can generate larger regions by iteratively filling in masked positions, controlling the length and global coherence of generated sequences typically requires additional heuristics~\citep{wang2019bert,ghazvininejad2019mask}.

For full, variable-length generation, autoregressive models offer an alternative approach. These models generate sequences one amino acid at a time from left to right, with each new residue conditioned on all previously generated residues. This enables open-ended sequence generation without predefined length constraints, making them well-suited for scaffold-free design (for example, ProGen and related AR models~\citep{madani2023large}). The trade-off is the loss of bidirectional context for mid-sequence redesign, which can be mitigated in practice by hybrid workflows that draft with AR decoding and then refine with span infilling. These conditioning mechanisms extend naturally beyond sequence context, which motivates explicit conditional generation on antigens, epitopes, or metadata.

\paragraph{Conditional generation.}
A natural transition from MLM-based redesign is that these models already condition on sequence context, so external conditions can be incorporated smoothly by augmenting the conditioning vector $c$ with target or metadata embeddings. To steer sequences toward specific targets, generators can be conditioned on antigen, epitope, or contextual embeddings such as HLA. This biases CDR regions or full receptors toward complementarity with a desired binding surface rather than sampling from generic repertoire statistics. In antibodies, IgLM~\citep{shuai2023iglm} exemplifies context-aware design: trained on 558\,M heavy- and light-chain sequences, it performs variable-length infilling such as CDR loops using bidirectional context and conditioning on species and chain type, enabling targeted library redesign with improved \textit{in silico} developability profiles. In T cells, TCR-TRANSLATE~\citep{karthikeyan2025conditional} and GRATCR~\citep{zhou2025gratcr} generate CDR3$\beta$ sequences conditioned on peptide–MHC context and can recover known binders for unseen epitopes, while CATCR-G~\citep{ji2024predicting} targets epitope-conditioned TCR design. Paired-chain antibody models such as p-IgGen~\citep{turnbull2024piggen} further support realistic heavy–light generation and provide a foundation for antigen-conditioned sampling as paired Ab–Ag datasets grow. Conditioning on continuous properties, for example, stability, solubility, or immunogenicity~\citep{li2021deepimmuno}, is less explored for antibodies and TCRs but is straightforward in principle, mirroring regression-guided conditional sequence design in small-molecule models~\citep{born2023regression, chang2024bidirectional}. 
Beyond discrete target/context embeddings, diffusion models can be property-conditioned via guidance: DIFFFORCE (diffusion differentiable force field) augments DDPM (denoising diffusion probabilistic model)~\citep{ho2020denoising} sampling with gradients from a differentiable force field to bias CDR generations toward lower interface energy~\citep{kulyte2024improving}. DDPM is a generative model that learns to create data by reversing a gradual noising process. Here it bias the energy at the contact surface between the antibody and antigen, where lower values indicate more stable binding, improving structural and sequence quality without retraining the prior. 

While conditioning enriches generation by steering models toward specific contexts or properties, it largely enforces constraints passively. In many applications, however, the goal is to actively optimize sequences with respect to multi-objective criteria rather than merely condition on them. This motivates reinforcement learning approaches that reshape the generator itself using reward signals derived from predictive or oracle models.

\paragraph{Reinforcement learning in sequence space.}
Sequence generators can be wrapped in reinforcement learning (RL), a machine learning model in which models learn to make decisions by receiving rewards for desired outcomes, to bias sampling toward sequences with higher predicted functional scores, such as affinity, stability, or developability. In practice, a generator, often a transformer, serves as the policy (the learned rule that decides which sequence to propose next) and is optimized with policy-gradient (a method for updating
the policy based on rewards) or offline-RL objectives against a composite reward built from predictive models and heuristics. For antibodies, AB-Gen combines a generative pre-trained transformer (GPT) policy with multi-objective rewards to design CDR3 of heavy chain (CDR-H3) libraries that pass stringent property filters~\citep{xu2023ab}. Stable online and offline RL for CDR-H3 design has also been demonstrated, with improved binding-energy proxies relative to Bayesian or evolutionary baselines~\citep{vogt2023stable}. Hybrid schemes connect RL with diffusion or flow backbones, for example RL-guided diffusion for antibody optimization, to balance exploration and constraint satisfaction~\citep{vogt2024betterbodies}. Closely related model-based approaches such as Conditioning-by-Adaptive-Sampling iteratively reweight a generative prior toward desired properties and have been applied to biological sequence design, offering an alternative to explicit RL while retaining oracle-guided improvement~\citep{brookes2019conditioning}. On the T-cell side, autoregressive TCR generators have been adapted with RL to skew generation toward peptide-specific recognition, demonstrating epitope-targeted sequence proposals after policy finetuning~\citep{lin2024tcr}. Yet many design goals concern not a single optimal sequence but a distribution over plausible CDR solutions, which motivates modeling full CDR ensembles with flow-based methods.

\paragraph{Flow-based models for CDR3 ensembles.}
Because CDR3 loops, especially antibody CDR-H3 and TCR CDR3$\beta$, exhibit high length and sequence variability, it is natural to treat their design as distribution modeling rather than single-point generation.  Normalizing flows are generative models that learn to transform a simple distribution (e.g., Gaussian noise) into a complex one (e.g., realistic CDR sequences) through a series of reversible steps, allowing both sampling of new sequences and calculation of their likelihood. In antibodies, recent work couples geometry and sequence explicitly: AntibodyFlow learns a 3D flow over loop geometry, for example distance matrices, and then conditions sequence generation on the sampled geometry, improving validity and geometric plausibility for designed CDRs~\citep{xu2024antibodyflow}; FlowDesign employs flow matching for sequence–structure co-design and reports gains in amino-acid recovery, RMSD, and Rosetta energy~\citep{wu2025flowdesign}; and IgFlow uses SE(3)-equivariant flow matching (a generative method whose predictions remain valid regardless of how the molecule is rotated or shifted in space) for \textit{de novo} variable-domain structure generation with a discrete-flow component for sequence~\citep{nagaraj2024igflow}. For TCRs, explicitly flow-based generators are only beginning to appear, but closely related diffusion models have been proposed for epitope-specific CDR3 generation, for example TCR-epiDiff~\citep{seo2025tcr}, suggesting a similar path for flow-based TCR ensemble modeling as paired TCR–epitope datasets expand. Methodologically, these approaches are enabled by recent advances that adapt flow matching to continuous trajectories and extend normalizing flows to discrete alphabets, which makes flow-based modeling feasible for categorical biological sequences~\citep{lipman2022flow,hoogeboom2019integer}.

\subsection{Clinical applications and case studies }
\begin{figure}[ht]
\centering
\includegraphics[width=0.9\textwidth]{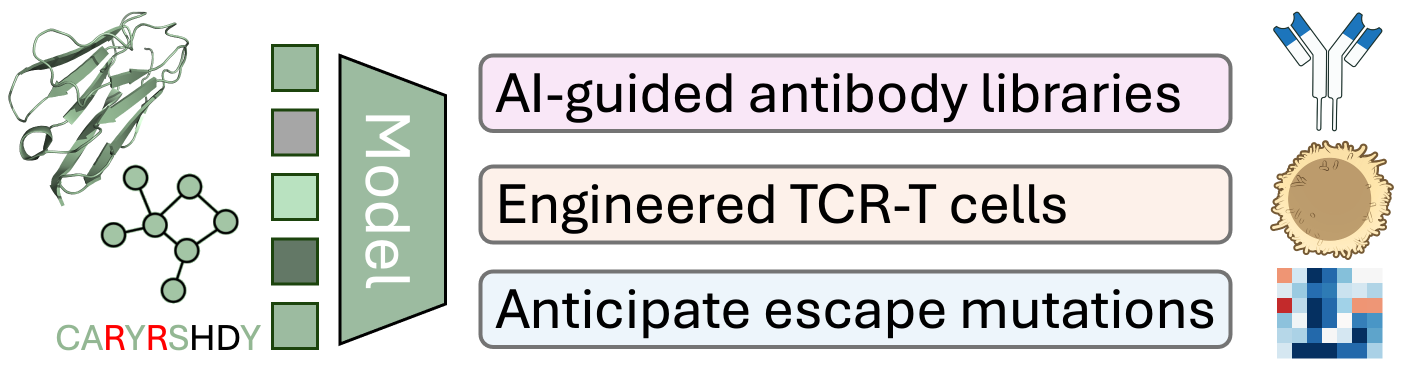}
\caption{Translational applications of AI models in immune receptor engineering. Generative and predictive models enable three major translational outcomes: (1) AI-guided antibody library design for therapeutic discovery, (2) engineered TCR-T cells for adoptive cell therapy targeting tumor neoantigens, and (3) anticipation of viral escape mutations to guide vaccine and therapeutic development.
}
\label{fig:applications}
\end{figure}

To bridge method to medicine (Fig.~\ref{fig:applications}), we highlight translational case studies where structure-native generative modeling has moved from in silico proposals to experimental validation and preclinical decision making. These examples pair diffusion and inverse folding generators with active learning, structural triage, and standardized assays, yielding candidates that meet clinical constraints on potency, safety, and manufacturability. We focus on three vignettes that capture this pipeline end to end: AI-guided antibody libraries, engineered TCR-T cells, and multi-objective optimization across efficacy and developability (Fig.~\ref{fig:applications}).

\paragraph{AI-guided antibody libraries.}
Recent work shows that generative models can deliver practical gains in discovery outcomes, not just new ways to design. Fine-tuned diffusion models have produced epitope-targeted VHHs and scFvs that validate as nanomolar binders against predefined surfaces when paired with yeast display, achieving targeted enrichment without loss of library diversity~\citep{bennett2025atomically}. Sequence-native generation focused on CDR-H3 has yielded functional SARS-CoV-2 binders directly from language models, demonstrating rapid retargeting to emergent variants~\citep{he2024novo}. Multi-step agentic workflows further concentrate wet-lab effort by proposing and triaging nanobody candidates end-to-end before synthesis and screening, which shifts experimental effort toward higher-quality subsets~\citep{swanson2025virtual}.

A concrete demonstration of library downselection and hit-rate gains comes from SARS-CoV-2 RBD campaigns (Wuhan, XBB.1, XBB.1.5): using AlphaFold3 and antibody-focused pipelines as priors and post hoc filters reduced a computational library of approximately 11,000 designs to a few hundred while preserving epitope coverage, and delivered double-digit yeast-display hit rates with successful re-epitoping and rescue~\citep{dreyer2025computational, abramson2024accurate}.

\paragraph{TCR-engineered T cells.}
Clinical use requires balancing affinity with safety, including off-target risk, cross-reactivity, and HLA restriction~\citep{golikova2024tcr}. Early clinical signals from personalized neoantigen products support this trajectory; BNT221 is an autologous, polyclonal T-cell product generated ex vivo by priming patient PBMCs with top-ranked MHCI and MHCII neoantigen peptides selected by a machine-learning pipeline that uses neural-network predictors of peptide–MHC binding, and it has shown preliminary activity in solid tumors~\citep{borgers2025personalized}. To meet safety constraints, systematic preclinical triage now integrates proteome-wide mimic searches, expression assays using mRNA and full-length proteins, and allo-reactivity screens; this revealed previously unrecognized HLA cross-recognition by the clinical 1G4 TCR and demonstrated tumor rejection by a CD20-specific TCR in an HLA-A02{:}01 transgenic model, establishing practical gates for IND-enabling evaluation~\citep{foldvari2023systematic}. This process is the regulatory application that must be approved before a new drug can be tested in humans.

Structure-based epitope-ranking tools such as TCRen leverage TCR–pMHC contact patterns to prioritize unseen epitopes for a given TCR, improving cognate-epitope identification and shortlisting many candidate neoepitopes for follow-up testing and in silico cross-reactivity triage~\citep{karnaukhov2024structure}. Likewise, conditional generative models have yielded experimentally validated receptors, including a WT1/HLA-A*02{:}01 TCR generated by TCR-TRANSLATE~\citep{karthikeyan2025conditional}.

Together, these approaches outline an integrated path where computation accelerates nomination and de-risks development, but ultimate success depends on rigorous safety gates, durable efficacy, and scalable manufacturing in the clinic~\citep{baulu2023tcr}.

\paragraph{Therapeutic design with generative models.}
Generative models are increasingly used to optimize therapeutic immune receptors such as antibodies and TCRs. By sampling novel sequences guided by learned fitness landscapes, these models can improve binding, reduce immunogenicity, and enhance developability.
In practice, therapeutic candidates must satisfy multiple design objectives at once—such as potency, specificity, stability, manufacturability, and human-likeness. Modern frameworks address this complexity by incorporating uncertainty-aware multi-task predictors and Pareto front selection strategies that jointly optimize competing objectives during the design phase~\citep{parkinson2025resp2}. Energy- and diffusion-based models further enable direct optimization of physicochemical and structural objectives during sampling, while developability and human-likeness filters help keep generated sequences within safe manifolds~\citep{cheng2024new, joubbi2024antibody}. When integrated with modern complex predictors such as AlphaFold3 for interface sanity checks and with wet-lab active learning, these workflows have begun to deliver antibody leads that balance binding with formulatability at clinically relevant thresholds.

In summary, generative AI offers a path from prediction to rational design, enabling the creation of novel antibodies and TCRs with improved therapeutic potential. Sequence-based and structure-based models address complementary aspects of the design problem: sequence models explore the vast combinatorial diversity of repertoires, while structure models enforce geometric constraints and biophysical plausibility. Their integration, particularly in combination with reinforcement learning and experimental feedback, is likely to transform the landscape of immune receptor engineering.

\section{Multi-modal and integrative approaches}

\begin{figure}[ht]
\centering
\includegraphics[width=0.9\textwidth]{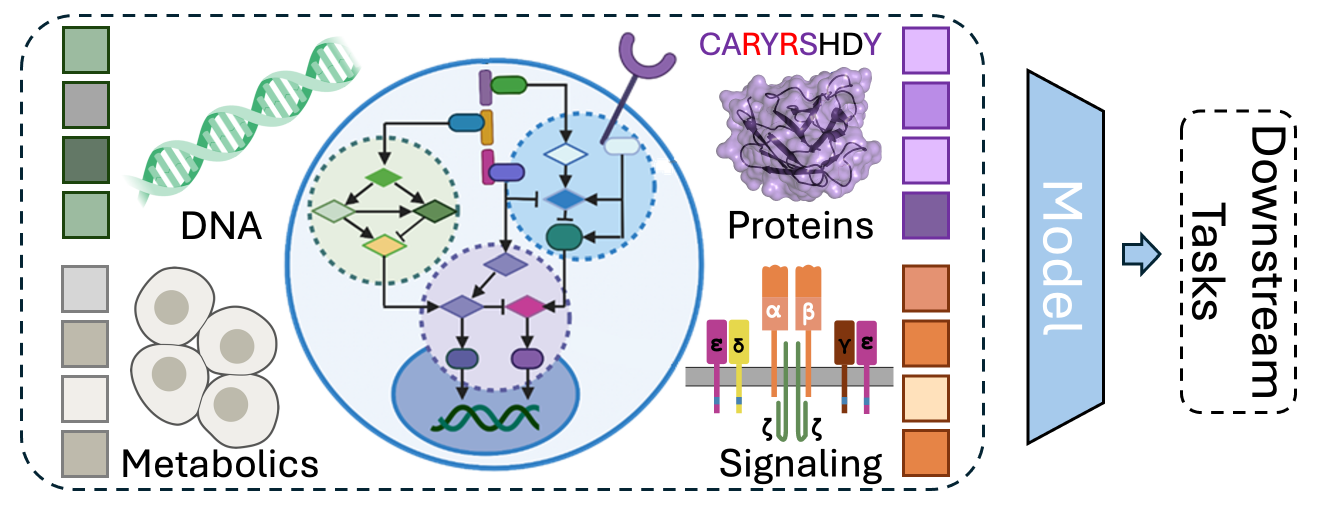}
\caption{Multimodal integration for immune receptor modeling. (Left) Diverse biological data modalities serve as inputs, including DNA sequences (green helix), metabolomic profiles (gray), protein sequences and structures (e.g., antibody sequence and 3D structure), and cellular state readouts capturing receptor context (e.g., TCR complex composition and activation signals measured by cytokine secretion and/or single-cell transcriptomics). Each modality is encoded into tokenized representations (colored squares). (Center) These modalities reflect coupled cellular processes—antigen recognition, signaling cascades, transcriptional programs, and metabolism—that together shape immune receptor function. (Right) A unified model integrates multimodal representations to support downstream tasks such as receptor specificity prediction, phenotype classification, binding affinity estimation, and therapeutic candidate prioritization.
}
\label{fig:multimodal}
\end{figure}

Immune receptor engineering increasingly requires models that can integrate diverse data modalities beyond raw sequences (Fig.~\ref{fig:multimodal}). TCR and BCR function is not determined by sequence alone but emerges from the interplay of receptor features, transcriptional states, antigen presentation, and host context. As illustrated in Fig.~\ref{fig:multimodal}, modern multimodal frameworks combine DNA-level information (germline genes, somatic mutations), protein-level features (sequence, structure, binding interfaces), metabolomic signatures, and cell signaling states (including the full TCR/BCR complex architecture) into unified representations. These multi-modal AI approaches attempt to unify these layers, providing more systematic and clinically relevant models of receptor behavior that capture the systems-level context in which immune receptors operate.

\subsection{Single-cell receptor and transcriptome integration}

Paired scRNA–VDJ (VDJ, referring to the variable gene segments that recombine to generate receptor diversity) datasets, combining single-cell transcriptomics (scRNA-seq) with matched immune receptor sequencing (VDJ), represent a recent advance in single-cell immunology. Improvements in capture chemistry and library preparation now allow the simultaneous recovery of full transcriptomes alongside TCR or BCR sequences at scale. This dual readout enables computational models to link immune receptor features with cellular phenotypes such as activation, exhaustion, or differentiation. Crucially, it also allows learning receptor specificity within the functional context of each cell, opening new avenues for modeling immune recognition in situ~\citep{liu2022spatial, mathew2021single}.

Early integrative methods, such as CoNGA~\citep{schattgen2022integrating} and Tessa~\citep{zhang2021mapping}, linked TCR sequence similarity to transcriptional neighborhoods or embedded both modalities to support joint clustering and phenotypic inference. Building on these foundations, newer deep learning approaches aim to construct shared latent spaces that unify receptor and transcriptomic information. For example, scNAT aligns scRNA-seq and scTCR-seq profiles into a joint embedding for downstream discovery~\citep{zhu2023scnat}; UniTCR learns cross-modal representations using contrastive learning~\citep{gao2024unified}; and mvTCR employs a multimodal variational autoencoder (VAE), a neural network that learns compressed representations while enabling generation of new samples, with expert fusion to capture antigen specificity while preserving cellular state~\citep{drost2024multi}. These approaches have improved the identification and characterization of antigen-specific T cell subpopulations and enabled mapping of clonal architectures to effector programs in disease~\citep{zhu2023scnat, gao2024unified, drost2024multi, schattgen2022integrating, gao2022single}.

For B cells, the Benisse framework integrates heavy and light chain sequences with single-cell gene expression to reveal activation gradients and repertoire–phenotype coupling~\citep{zhang2022interpreting}. These analyses uncovered receptor–state associations that would be missed by sequence data alone, including enhanced coupling between BCR repertoires and transcriptomic states in the context of COVID-19~\citep{zhang2022interpreting}. Complementary efforts have assembled paired scRNA–BCR datasets labeled for antigen specificity and benchmarked models trained on expression features, repertoire features, and their combination. In these comparisons, models leveraging expression data consistently outperformed sequence-only baselines, highlighting the added value of integrating both modalities~\citep{erlach2024evaluating}.

\subsection{Multi-omics integration}

Beyond receptor–transcriptome profiling, integrative approaches increasingly incorporate additional omics layers to enhance predictive modeling. Notable additions include RNA–protein co-assays from techniques that simultaneously measure RNA and surface proteins in single cells (e.g., CITE-seq or REAP-seq), chromatin accessibility profiles from scATAC-seq (single-cell Assay for Transposase-Accessible Chromatin sequencing, which maps open chromatin regions), metabolomic signatures, spatial transcriptomics, immunopeptidomics, and perturbation readouts from CRISPR screens. These modalities often capture complementary cellular signals that are not accessible through transcriptomic or sequence data alone~\citep{furtwangler2025mapping}.

On the protein modality side, scmFormer integrates transcriptomic and single-cell proteomic data to enable cross-modal prediction and label transfer, offering a blueprint for RNA–proteome fusion at single-cell resolution~\citep{xu2024scmformer}. Focusing specifically on RNA and targeted surface proteins, totalVI jointly models CITE-seq counts within a shared latent space, enabling denoising and cross-modal inference. This approach improves marker recovery and allows protein imputation from RNA, refining antigen-specific cell identification compared to transcriptome-only models~\citep{gayoso2021joint}.
Extending to chromatin modalities, MultiVI integrates scRNA-seq with scATAC-seq within a unified probabilistic framework, enabling cross-modal inference even when one modality is missing~\citep{ashuach2023multivi}. By predicting chromatin accessibility from transcriptional state, it improves cell clustering and annotation while uncovering activation-associated regulatory programs.

For antigen specificity, barcoded antigen screening technologies offer direct supervision that can be integrated with receptor sequence features. \textit{LIBRA-seq} links BCR sequences to antigen-binding profiles at single-cell resolution, enabling prioritization of binders and down-selection of candidate antibodies at scale~\citep{setliff2019high}. Training on multi-antigen panels, this approach accelerates the discovery of broadly neutralizing antibodies by learning from antigen–receptor binding patterns across diverse epitopes.

At the epitope and presentation layer, immunopeptidomics-driven models combine mass spectrometry–identified ligands with peptide sequence and sample-level covariates to predict peptide–HLA presentation and prioritize neoepitopes. Representative frameworks include major histocompatibility complex analysis with recurrent integrated architecture (MARIA), which integrates peptide sequence, expression, and antigen processing features~\citep{chen2019predicting}; ImmuneApp, which leverages large ligand datasets for improved epitope ranking and motif deconvolution~\citep{xu2024immuneapp}; and MIX-TPI, which incorporates physicochemical descriptors of the pMHC complex to enhance generalization across HLA alleles and sample contexts~\citep{yang2023mix}. In translational settings, these predictions help narrow candidate targets in patient samples and focus downstream TCR screening on peptides most likely to be presented.

\subsection{Transfer learning across biological domains}

Data scarcity, particularly for paired receptor–ligand observations, remains a major challenge in immune receptor modeling. A powerful strategy to overcome limited labels is transfer learning~\citep{pan2009survey}, an approach where knowledge learned from one task or dataset is applied to improve performance on a related but different task, where models are first pretrained on large, general datasets to learn broad statistical patterns, and then fine-tuned for more specialized downstream tasks. For sequence-based immune modeling, a prominent example of transfer learning is the use of PLMs \cite{rives2021biological, elnaggar2021prottrans, madani2023large}. 

These models are trained on large collections of non-immune protein sequences and learn statistical patterns that reflect biochemical properties, evolutionary constraints, and approximate structural features~\citep{lin2022language, burbach2024improving}. In recent years, PLMs specifically trained on BCR \cite{leem2022deciphering, olsen2022ablang, shuai2023iglm} or TCR \cite{wu2021tcr, myronov2023bertrand} repertoires have also emerged, offering specialized representations of immune-specific patterns. Both generalist and immune-specific PLMs can be adapted for downstream immune tasks, such as paratope prediction, affinity regression, or receptor–ligand scoring. By leveraging these pretrained priors, PLM-based models improve generalization and can achieve good performance with fewer labeled examples, particularly in settings where data are scarce~\citep{schmirler2024fine, deutschmann2024domain}.

PLM-derived representations can also be used in models that handle incomplete observations. For example, TCR datasets often provide only $\beta$-chain information, lacking full ($\alpha$ \textbar{} $\beta$ \textbar{} peptide) triplets.
In this setting, multimodal masked objectives and modality dropout can still support learning and inference with incomplete input views; TULIP exemplifies this strategy and improves robustness under $\beta$-only supervision~\citep{meynard2024tulip}. 

More broadly, transfer learning is most effective when the data and objectives used for pretraining are similar to those of the downstream task, so that the learned representations remain relevant. When this holds, adapting general-purpose PLMs to immune receptor tasks can improve performance, even when receptor–ligand data are scarce or incomplete.

To summarize, multi-modal and integrative approaches represent a pivotal advance in modeling immune receptor function. By linking receptor sequences with gene expression, phenotypes, and other omics layers—such as chromatin accessibility, proteomics, and immunopeptidomics—AI models can move beyond isolated prediction tasks to capture the complexity of adaptive immunity. In the long term, such integrative frameworks hold promise for personalized immune engineering, where receptor design is tailored not only to the target antigen but also to individual transcriptional profiles, HLA genotypes, and the surrounding tissue or tumor microenvironment.

\section{Key limitations and open challenges}
The rapid progress of artificial intelligence in immune receptor engineering has opened new possibilities for the design and optimization of therapeutic antibodies, TCRs, and vaccines. However, realizing the full potential of these advances requires overcoming persistent challenges across multiple fronts. Some of the main challenges are data-related constraints that hinder robust model training, model-specific challenges that compromise prediction reliability and interpretability, and translational bottlenecks that impede the integration of computational outputs into experimental and clinical pipelines.

\paragraph{Data-related challenges} A significant challenge in this field is the limited availability of data. Most datasets are heavily biased toward viral epitopes presented by just a few common HLA alleles, leaving the vast diversity of HLA types underrepresented. As a result, models trained on this skewed data often struggle to generalize to the broader range of antigens encountered in real-world clinical settings~\citep{weber2024t}. This issue is exacerbated by the lack of true negatives, which are critical for building robust machine learning models. In the absence of experimentally confirmed negative examples, researchers often create them artificially, i.e. by using random shuffling or decoy sequences, which can introduce systematic bias. In addition, inconsistencies in experimental protocols and the scarcity of longitudinal data on clonal dynamics limit the ability to integrate datasets or model immune responses over time.

Beyond issues of data bias and scarcity, deeper computational bottlenecks arise from experimental limitations. Key challenges include the lack of paired-chain information, uneven sampling across immune repertoires, and inconsistent protocols across studies. Although the volume of immune receptor data continues to grow, these gaps limit the quality and usability of the data for training robust models. In particular, the shortage of high-quality paired receptor–ligand structures, such as antibody–antigen or TCR–pMHC complexes, constrains the development of supervised, structure-aware models that require labeled examples to learn accurate binding interactions.

A related challenge lies in integrating multiple modalities. Modern datasets often include sequences, structural models, repertoire frequencies, and phenotypic measurements. Each modality captures a different aspect of receptor biology, yet combining them into coherent models is technically complex. Effective integration requires algorithms capable of handling data at disparate scales and resolutions, while also managing variable noise levels, all without losing the biological meaning embedded in each layer.

Finally, longitudinal datasets tracking clonal dynamics over time are also rare, making it difficult to model the evolution of immune responses or integrate data across time points and individuals. Together, these limitations in experimental design, data quality, and coverage represent foundational obstacles to progress in AI-driven immune receptor modeling.

\paragraph{Model-related challenges} A second major challenge is the lack of interpretability in most current models. Deep learning approaches, especially transformer-based protein language models, can produce accurate predictions, but they often behave as ``black boxes," offering little insight into the underlying mechanisms of immune recognition. Furthermore, most models provide only local explanations, i.e. explanations that only hold for individual examples, rather than uncovering broader principles that govern TCR–pMHC interactions. Recent work has begun to address this gap by using attention-based architectures and post-hoc rule-extraction pipelines to highlight informative residues and sequence motifs~\citep{weber2021titan,papadopoulou2022decode, weber2024t, chen2024tepcam, zhang2024protein}, but these methods still provide only partial insight and require further development.

In addition, their confidence scores are often poorly calibrated, which can mislead clinical decision-making. Without proper uncertainty estimation, it becomes difficult to distinguish between model error and intrinsic data noise~\citep{lambert2024trustworthy}. These issues are compounded by poor generalization: models frequently fail on peptides that differ from those seen during training, with performance dropping sharply on out-of-distribution examples~\citep{castorina2025assessing}. This lack of robustness raises serious concerns about their reliability in therapeutic applications.

Compounding these issues are deeper structural and biophysical challenges that many models fail to address. A central limitation lies in their inability to capture the dynamic nature of receptor–antigen binding. For instance, the complementarity-determining region 3 (CDR3) loops, critical for antigen recognition, are highly flexible, adopting different conformations depending on the binding target and the context. Most structure predictors treat these loops as static, leading to inaccurate representations of the interaction interface. As a result, structural uncertainty further limits the reliability of downstream predictions.

Data scarcity and model limitations together constrain the ability of current approaches to make robust, transferable predictions. Models trained on receptors from a specific pathogen, tumor type, or patient cohort often perform poorly when applied to new contexts. This lack of transferability reflects the underlying heterogeneity of immune systems, shaped by individual immune histories and diverse HLA backgrounds. As a result, even high-performing models can fail when confronted with novel repertoires or antigen landscapes. Bridging this gap requires not only larger and more representative datasets, but also new modeling frameworks that are robust to structural flexibility, multimodal complexity, data imbalance, and population-level diversity. Generalization across datasets and patients remains one of the most difficult open problems in computational immunology.

\paragraph{Translational challenges: } A third major challenge lies in the translational gap between computational predictions and clinical application. Experimental validation remains slow and resource-intensive, creating a bottleneck in which models can generate thousands of candidate receptors, but only a small subset can feasibly be tested in the lab. The field continues to lack standardized benchmarks for comparing model performance and robust frameworks for integrating BCR and TCR binding predictions into therapeutic development pipelines. 

Importantly, predicted binding affinity does not equate to therapeutic efficacy. Receptor binding is necessary but not sufficient to trigger an appropriate immune response; activation, signaling strength, and cellular context also play essential roles. In addition, therapeutic candidates must satisfy requirements beyond binding, such as pharmacokinetic stability and low off‑target toxicity, that current models rarely evaluate. These scientific gaps are compounded by practical ones: regulatory pathways for AI‑designed immune therapeutics remain poorly defined, and reproducibility issues across studies continue to slow progress toward clinical translation.

\section{Future directions}

Despite these challenges, several promising directions are emerging at the intersection of computational modeling and immunotherapy. The success of protein language models highlights the potential for foundation models trained specifically on immune receptor data. Such models could unify diverse tasks, ranging from specificity prediction and affinity regression to generative design, within a single framework. At sufficient scale, immune-focused foundation models may enable zero-shot predictions of binding or cross-reactivity, reducing reliance on task-specific labeled datasets. Early efforts using transfer learning from general protein models have already shown encouraging results, pointing toward the need for dedicated models tailored to adaptive immunity.

At the same time, there is a central opportunity to make fuller use of structural information. For proteins broadly, accurate 3D structures provide crucial information about binding interfaces, conformational transitions, and allosteric regulation, all of them key elements that bridge the gap between sequence-level patterns and functional outcomes. In the context of immune receptors such as TCRs and antibodies, structural information is even more critical because the CDR loops, especially the highly variable CDR3 loops, are inherently flexible and often undergo substantial conformational rearrangements upon antigen binding. Yet, most current structural predictors \cite{abramson2024accurate, chai2024chai, passaro2025boltz} still rely on static snapshots that capture only a limited view of this flexibility, failing to account for ensemble behaviors and induced-fit dynamics that are central to binding and activation.

A promising direction, therefore, is to integrate AI with biophysical simulations to capture immune receptor flexibility better. Physics-informed generative models, guided by molecular dynamics (MD) simulations, offer a framework for sampling conformational ensembles that respect both physical laws and empirical data. By accounting for conformational plasticity, these hybrid approaches could improve predictions not only of binding affinity but also of downstream functional outcomes such as receptor activation and signaling potential. Combining AlphaFold-style structure predictors with MD-derived conformational ensembles may ultimately yield models that more accurately reflect the dynamic landscape of receptor–ligand interactions.

To fully harness such models in clinical contexts, however, interpretability remains essential. Black-box predictions are insufficient when therapeutic safety is at stake. Recent advances, including attention maps that highlight key interface residues, residue-level saliency methods, and causal inference frameworks, aim to improve mechanistic insight and reliability. By making AI models more transparent, these approaches can help build trust among experimentalists and clinicians while also advancing our understanding of immune recognition.

Interpretability alone is not enough. Real-world deployment of AI-designed immune therapeutics will require tight integration with experimental pipelines. Iterative cycles of in silico design, high-throughput screening, and wet-lab validation are needed to ensure that predictions translate into functional and safe therapeutics. Concurrently, regulatory frameworks must evolve to evaluate these new classes of molecules. Establishing standards for benchmarking, reproducibility, and uncertainty quantification will be critical to ensure robust, clinically actionable outputs.

Artificial intelligence is thus emerging as a transformative force in immune receptor engineering, accelerating the shift from descriptive to design-driven science. While antibody modeling is already shaping drug discovery pipelines, TCR modeling, though less mature, holds great promise for personalized immunotherapies. Looking ahead, the convergence of physics-informed AI, immune-specific foundation models, and interpretable design strategies will empower researchers to engineer receptors that are not only high-affinity and specific but also safe, stable, and personalized. The long-term vision is a scalable, data-driven framework that integrates computational prediction with experimental validation to enable the next generation of immunotherapies and vaccines.

\bibliography{main}
\end{document}